\documentclass{rmaa}

\title{Very Large Array Observations  of Winds from Massive Stars}
\shorttitle{Ionized Winds of Massive Stars}
\shortauthor{Contreras {\it et al.}}
\author{M.~E.~Contreras,\altaffilmark{1} 
G.~Montes,\altaffilmark{1,2}
F.~P.~Wilkin\altaffilmark{1}
}

\altaffiltext{1}{Centro de Radioastronom\'{\i}a y Astrof\'{\i}sica, UNAM}
\altaffiltext{2}{Universidad Michoacana de San Nicol\'as de Hidalgo, Morelia, Michoac\'an, M\'exico}

\fulladdresses{
\item M.~E.~Contreras, G.~Montes \& F.~P.~Wilkin:
Centro de Radioastronom\'{\i}a y Astrof\'{\i}sica,
UNAM Campus Morelia, Apdo.Postal 3-72 (Xangari), 58089 Morelia, Michoac\'an,
M\'exico (m.contreras@astrosmo.unam.mx).
}

\resumen {El modelo est\'andar para la emisi\'on libre-libre de vientos ionizados se basa 
en
la suposici\'on de un viento estacionario, isotr\'opico y homog\'eneo. Sin embargo, durante
la \'ultima d\'ecada estas suposiciones han sido cuestionadas. Existen objetos cuyo viento
no se comporta como lo predice el modelo est\'andar.
En este trabajo presentamos resultados para tres fuentes brillantes: P Cyg, 
Cyg OB2 No.~12 y WR 147. Se ha reportado que estos objetos poseen vientos que
se desv\'\i an de lo esperado. Reportamos el flujo observado y derivamos el 
tama\~no, el \'\i ndice espectral y la tasa de p\'erdida de masa para cada uno de ellos. 
Estos par\'ametros nos permiten detectar la presencia de posibles asimetr\'\i as, 
inhomogeneidades y variaciones en la densidad de flujo. Estas desviaciones confirman 
un comportamiento distinto al cl\'asico en estos vientos.}

\abstract {The classical model for free-free emission from ionized stellar
winds is based on the assumption of a stationary, isotropic and homogeneous wind.
However, since there exist objects whose wind behaviour deviates from the standard model,
during the last decade these assumptions have been questioned. In this work, we present
results for 3 bright sources: P Cyg, Cyg OB2 No.~12 and WR 147. These objects have been 
reported to possess winds that deviate from the basic assumptions. We have obtained flux 
densities, sizes, spectral indices and mass loss rates for each of the targets. These 
parameters allow us to analyze possible asymmetries, inhomogeneities and time variations 
in the flux densities. These features confirm the nonclassical behaviour of these
winds.}


\begin{document}
\maketitle


\section{Introduction}

 Since winds from massive stars are ionized by the ultraviolet radiation from the 
underlying star, they can be studied observationally at radio wavelengths through their 
free-free emission. Until recently, massive star winds were treated as isotropic and 
homogeneous flows with constant velocity and known electron temperature, degree of 
ionization and chemical composition. In this way, based on the classic work of 
Panagia \& Felli (1975) and Wright \& Barlow (1975), it was possible to obtain a value 
for the mass loss rate $\dot M$ from radio observations. However, since the work of 
Abbott, Bieging \& Churchwell (1981), who found that in some cases these assumptions 
are not valid, various authors have found firm evidence for deviations from these basic 
assumptions.

\begin{table*}[!t]
\begin{center}
\renewcommand{\footnoterule}{\rule{0cm}{0cm}}
\caption{Source Sizes and Fluxes}
\vskip 0.8cm
\begin{tabular}{lcccccc}
\hline \\[-1ex]
Source & $\theta_{0.7cm}$(PA) & $\theta_{3.6 cm}$(PA) & $\theta_{6 cm}$(PA) & $S_{0.7cm}$ & $S_{3.6cm}$ & $S_{6cm}$ \\
& [arcsec][deg] & [arcsec][deg] & [arcsec][deg] & [mJy] & [mJy] & [mJy] \\[2ex]
\hline \\
P Cyg & $0.06 \times 0.03$ (42) & $0.19 \times 0.13$(160) & $0.22 \times 0.19$(29)  
                        & 9.7 $\pm 2.1$ & 8.0 $\pm 0.2$ & 6.0 $\pm 0.1$ \\[0.7ex]
Cyg OB2 No.~12 & $0.04 \times 0.03$(32) & $0.13 \times 0.12$(103) & $0.24 \times 0.12$(176) 
                        & 9.0 $\pm 1.5$ & 5.9 $\pm 0.1$  & 4.2 $\pm 0.1$ \\[0.7ex]
WR 147 S& 0.06 $\times$ 0.05(22) & 0.25 $\times$0.21(120) & 0.46 $\times$0.35(163)  
                        & 36.6 $\pm 2.2$ & 24.5 $\pm 0.1$ & 22.1 $\pm 0.2$ \\[0.7ex]
WR 147 N & $\cdots$ & 0.37 $\times$ 0.30(110) & 0.49 $\times$ 0.21(93) 
                        & $\cdots$ & 9.3 $\pm 0.2$ & 10.8 $\pm 0.2$ \\[0.7ex]
\hline
\vspace{-0.7cm}
\end{tabular}
\end{center}
\vspace{0.5cm}
{{\scshape Note}.$-$ Deconvolved size errors are 
$\sim 0{\rlap.}{''}01$.}
\end{table*}

 The existence of possible inhomogeneities has been studied, both observationally 
and theoretically. 
Radio observations seem to detect (marginally) blobs of material moving with the wind 
(Skinner et al.~1998; Contreras \& Rodr\'\i guez 1999; Exter et al.~2002). 
In the optical region, peaks superposed on 
emission lines are suggested to represent individual blobs (Moffat \& Robert 1994). 
Theoretically, Cherepashchuck (1990) studied the discrepancy between the 
expected and the observed X-ray luminosity. He concluded that a model 
including inhomogeneities can explain this discrepancy and suggests that as much as 
80\% of the mass of the wind could be in the form of blobs. In fact, Williams 
et al. (1997) reported firm evidence of an asymmetric wind in WR 147, making the WR 
star of this binary system a very good example of a non-classical wind source.

  Variable radio emission has been detected from several massive stars
including P Cyg (Abbott, Bieging \& Churchwell 1981; Contreras et al.~1996; 
Williams et al.~1997). This variability suggests that we are not dealing with
a stationary wind. It has been proposed that these variations are
due to a changing mass loss rate. However, the variability in the observed
optical emission lines can be explained by the presence of an asymmetric
stellar wind in rotation. In the case of a binary system, where there is an interaction
surface between the two stellar winds, the variability can be due to 
the orbital sweeping of this interaction surface even in the absence of intrinsic
variability of the stellar winds (Girard \& Willson 1987; Luehrs 1997; 
Georgiev \& Koenigsberger 2002).

  The study of the main assumptions of the classical wind model is very
important because mass loss rate values are usually derived assuming a
classical thermal wind. Thus, in order to determine reliable mass loss rates, 
we should take into account possible deviations (hacer algo en cada fuente). 
In this paper, we present new VLA observations at 0.7, 3.5 and 6 cm of three bright 
radio sources: P Cyg, Cyg OB2 No.~12 and WR 147.

\section{Observations}

\begin{figure*}[!t]\centering
\begin{center}
\centerline{\includegraphics[width=23pc,height=38pc,angle=270]{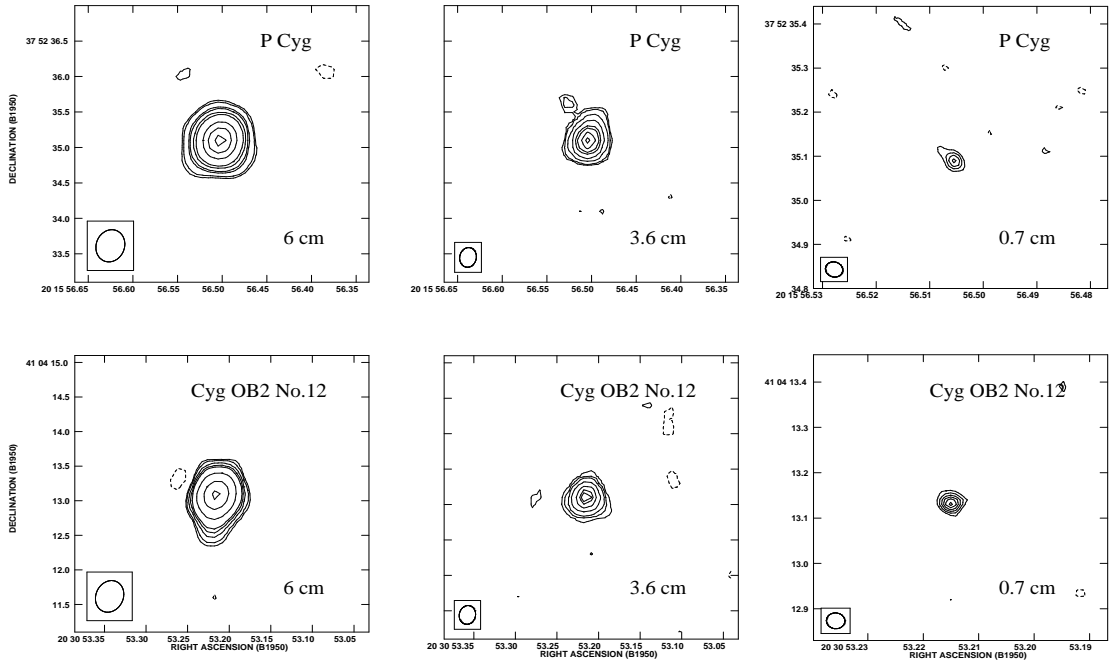}}
\caption{VLA maps of P Cyg and Cyg OB2 No.~12 at all observed wavelengths. All maps
were obtained using AIPS routine IMAGR with an intermediate data weight (ROBUST=0).
In the 3.6 cm map of P Cyg we can see a possible blob to the N-E in the wind, as well
as one in the Cyg OB2 No.~12 map at 6 cm, to the S-E. Both weak emission regions
are $\sim3-4\sigma$ above the map noise of $6.48\times10^{-5}$ (P Cyg) and 
$5.72\times10^{-5}$ (Cyg OB2 No.~12) at 3.6 and 6 cm, respectively.}
\end{center}
\end{figure*}

  We present two sets of observations taken with the Very Large Array (VLA) of the 
NRAO\footnote{The National Radio Astronomy Observatory is operated by Associated 
Universities Inc. under cooperative agreement with the National Science Foundation}
on 1999 June 28 and September 3 at 0.7, 3.6 and 6-cm. During both epochs the 
array was in the ``A'' configuration giving angular resolutions of 
$\sim 0{\rlap.}{''}05$, $\sim 0{\rlap.}{''}2$ and $\sim 0{\rlap.}{''}4$ for
each wavelength, respectively. During the first observing run all three sources were
observed, while during the second only WR 147 was observed. The total on-source 
integration time was $\sim$ 1 hour for each source for the first run and 1.5 hours for 
the second. Amplitude calibrators were 1328+307 and 0137+331 and phase calibrators 
were 2005+403 and 2015+371 on each observing run. For the first season, the observed 
bootstrapped flux densities for 2005+403 were: 1.18$\pm 0.06$, 2.01$\pm 0.02$ and 
2.54$\pm 0.03$ mJy at 0.7, 3.6 and 6 cm respectively and for 2015+371: 1.35$\pm 0.09$
at 6 cm. For the second run only 2005+403 was observed and a flux density of 
1.99$\pm 0.01$ at 3.6 cm was obtained.
 
  Data reduction was performed using the Astronomical Image Processing System (AIPS) 
software of the NRAO. At 3.5 and 6 cm we have followed standard VLA procedures for 
editing, calibrating and imaging. At 0.7 cm, we have applied a small correction to the 
data for atmospheric extinction using a zenithal optical depth of $\tau$ = 0.1. In this 
way, we presume to have minimized the expected tropospheric opacity dependence of the 
aperture efficiency.

  CLEANed maps were obtained using different data weighting schemes according to the
source. For P Cyg and Cyg OB2 No.~12, an intermediate weight between Natural and Uniform
was used (ROBUST=0, Figures~1 and 2). While for WR 147, pure uniform 
weighting was preferred (Figure~3). In order to obtain flux densities and 
source sizes, we have used the AIPS IMFIT procedure. Based on a 2D-Gaussian 
fit to the source, we have determined integrated flux densities (Table 1) 
and deconvolved sizes (Table 2) for all sources. Distances and wind terminal
velocity values were taken from the literature (Contreras et al.~1996); these
data, together with the relation derived by Panagia \& Felli (1975) and 
Wright \& Barlow (1975), allowed us to derived mass-loss rates for each object
(Table 2). Our values are not corrected for the effects of possible asymmetries or
inhomogeneities, but they are expected to be correct to an order of magnitude.

\begin{figure*}[!t]
\begin{center}
\centerline{\includegraphics[width=13pc,height=6pc,angle=270]{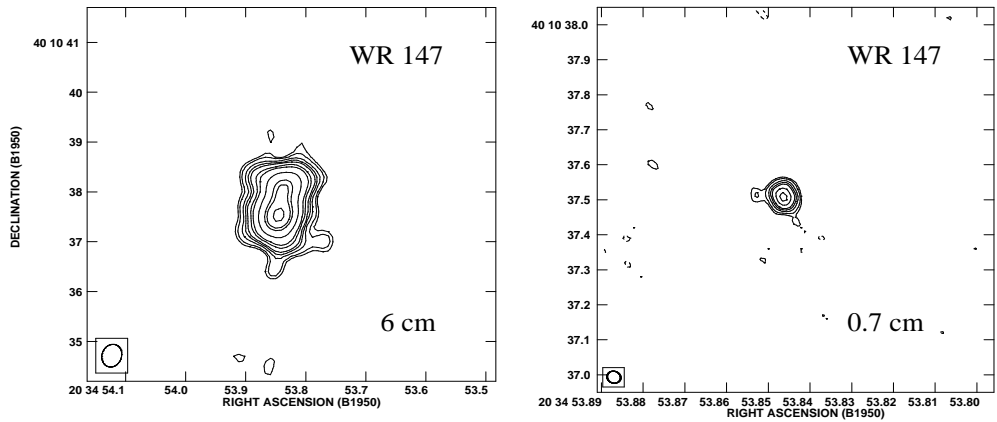}}
\caption{VLA CLEANed maps of WR 147 at 0.7 and 6 cm. Maps were obtained using AIPS
routine IMAGR with an intermediate data weight (ROBUST=0). In the 6 cm map we can 
see two faint small emitting regions to the south of the WR star.}
\end{center}
\end{figure*}

\begin{table}[!b]
\begin{center}
\renewcommand{\footnoterule}{\rule{0cm}{0cm}}
\caption{Derived Parameters}
\begin{tabular}{lccccc}
\hline \\[-1ex]
Source & $\alpha_{0.7-3.6cm}$ & $\alpha_{3.6-6cm}$ & $\dot{M}_{0.7cm}$ \\
& & &[10$^{-5} M_{\odot}$] \\[2ex]
\hline \\
P Cyg & 0.12 $\pm 0.13$ & 0.50 $\pm 0.05$ & 0.5$\pm$0.2 \\[0.7ex]
Cyg OB2 No.~12 & 0.26 $\pm 0.10$ & 0.63 $\pm 0.06$ & 2.7$\pm$1.0 \\[0.7ex]
WR 147 S & 0.25 $\pm 0.04$ & 0.18 $\pm 0.02$ & 1.0$\pm$0.4 \\[0.7ex]
WR 147 N & $\cdots$ & -0.25 $\pm 0.04$ & $\cdots$  \\[0.7ex]
\hline
\end{tabular}
\end{center}
\end{table}

\section{Results}

\subsection{P Cyg}

This B1 Ia star has been studied in the radio since the early 1980's when Abbott, 
Bieging \& Churchwell (1981) reported for the first time variable radio emission.
Radio emission variations are as large as $\sim$50\% without considerable
changes in the spectral indices (Contreras et al.~1996). Several explanations related 
to wind parameters (ionization degree, wind velocity, mass-loss rate) have been 
proposed, but none of them seems to fit the observed behaviour. In this work  we have 
determined flux densities at 0.7, 3.6 and 6 cm, based on our 2D Gaussian fits. 
Comparing these new flux densities with previous ones, we found that they 
have decreased by $\sim$70\%, $\sim$35\% and $\sim$34\%, respectively, from their 
1995 values (Contreras et al. 1996). This kind of behaviour has been observed by
Skinner et al. (1998) in their long-term deep radio study. They found that P Cyg shows 
important changes in its flux density where variations of 20-50\% are frequently
observed.

Regarding spectral indices, Skinner et al. (1998) found two different values 
corresponding to two distinct regions around P Cyg: the inner region (about $10''$ 
in radius) presents a spectral index of 0.8 while the outer region of the nebula shows 
a spectral index close to zero. In our case the spectral index behaviour
seems to be the opposite: the 0.7-3.6 cm spectral index, coming from a region closer
to the star, is almost flat (Table 2) while the 3.6-6 cm spectral index, coming from
regions farther out from the star, is consistent with the 0.6 value for a thermal wind.
The 0.7-3.6 cm spectral index has been altered by the decrease of the 0.7 cm flux 
density. As Skinner et al. (1998) proposed, since different wavelengths sample different
regions in the wind, each one with a completely different time-scale variation, one can 
expect that flux density changes may be poorly correlated or entirely uncorrelated. 
Then, we are observing a rapid change in the wind emission at 0.7 cm ocurring in the 
inner region but not observed in the outer regions of the wind.
Wind variability has been observed at other wavelengths. Recently, 
long-term optical photometric studies by van Genderen, Sterken \& De Groot (2002) and 
de Jager (2000) have described various types of stellar oscillations that P Cyg undergoes
which could be related to changes in the wind parameters. Thus, in light 
of these new studies, it would be useful to examine again the proposed explanations
mentioned above. 

Regarding morphology, we present 0.7, 3.6 and 6 cm maps shown in Figure 1. 
From these maps, P Cyg seems to have a non-isotropic wind at all three wavelengths
being more obviously elongated at 0.7 cm where its major/minor axis ratio is 2 
(Table 2). The observed wind asymmetry is consistent with the results of Skinner 
et al. (1998), who found a slightly asymmetric structure surrounding the compact
stellar wind source. Moreover, they found that the nebula around P Cyg is very
inhomogeneous or filamentary at various scales: a compact core of a few tenths of
an arcsec in diameter, an inner nebula of $\sim16-18''$ in diameter and an outer
nebula extending to $90''$. Although, in our 0.7 and 6 cm maps we can hardly see 
what could be emission from a blob denser than the surrounding wind, evidence of
a blob is more obvious in our 3.6 cm map. Here we can see emission at a 
$4\sigma$-level of a small condensation to the NE of P Cyg. This inhomogeneity is 
about $0\rlap{.}{''}6$ away from the central star which indicates that at an 
intermediate scale between the central core and the inner nebula (as defined by 
Skinner et al.) the wind is also inhomogeneous. 

 The derived mass-loss rate for P Cyg, while somewhat lower than that reported by 
Contreras et al.~(1996), is still consistent within the error, and consistent with
the value of $1-3\times 10^{-5}$ M$_\odot$ yr$^{-1}$ reported in the literature
(Skinner et al. 1998). Since our two values were obtained using the 0.7 cm flux 
density, to avoid possible non-thermal emission contamination (which showed a very 
important decrease in 1999), it is not surprising that the 1999 mass-loss rate is 
lower than the previous one. P Cyg's wind and associated nebula, require more
detailed study because the radio variability has yet to be explained.

\subsection{Cyg OB2 No.~12}

 Although no variable radio emission has been reported for this source, our new 
VLA observations show some variation in its 0.7, 3.6 and 6 cm flux 
densities. Both, 0.7 and 3.6 cm fluxes have decreased by $\sim$60\% and $\sim$18\%,
respectively, while its 6 cm flux has increased by $\sim$12\%. Our flux densities were
obtained from a 2D Gaussian fit to the source. These fluxes are shown in Table 1.
The flux variations have caused the spectral indices to change as well. The 3.6-6 cm 
index has decreased from those reported by Contreras et al.~(1996), becoming more 
consistent with the classical value for a thermal wind of 0.6, while the 0.7-3.6 cm 
index has decreased to a rather flat value of 0.3$\pm$0.1. 

Maps at the three observing wavelengths are shown in Figures 1 and 2. Based on its 
morphology, Cyg OB2 No.~12 does not show any deviation from a spherical wind at
0.7 and 3.6 cm  (Figures 1 and 2), however, at 6 cm its shape is obviously elongated 
(Figure 1). From a 2D Gaussian fit, we have obtained its angular 
size at all three wavelengths; it is clear that at 6 cm, the source has a major to 
minor axis ratio of 2 (Table 1). The contour map shows that the 6 cm emission does not
have a smooth oval shape but rather shows a kind of enhaced emission region to the south. 
Furthermore, if the source had an asymmetric wind, its shape would be elongated at all 
wavelengths, which is not the case. Since free-free emission depends on density
as $S \propto n_e^2$, we propose that this protuberance could be explained as a blob 
denser than its surroundings. Thus, the wind of Cyg OB2 No.~12 deviates from the 
standard assumption of a homogeneous wind.

\subsection{WR 147}

This very interesting system is now considered the archetype of colliding wind 
binaries. It clearly shows the presence of a wind interaction region curving away
from the Wolf-Rayet(WR) star, and emitting non-thermal radiation (north radio 
component). Its southern radio component (WR star) shows time-variable radio emission 
as well as inhomogeneities (Contreras \& Rodr\'\i guez 1999; Watson et al. 2002). 
In this work, we present new maps obtained from our 0.7, 3.6 and 6 cm data (Figures 2 
and 3). In these maps we can see both components and some of their fine structure. 

Flux density variability has been reported for both radio components (Williams
et al.~1997; Setia Gunawan et al.~2001). Although Setia Gunawan et al. (2001) find
variability at 6 and 21 cm for the north component of the system, from our new values 
(Table 1), we cannot see any significant variation in the 3.6 cm flux density of 
WR 147N, compared to its 1996 values. Thus, its 3.6 cm flux density may have remained 
constant, within the error, over an interval of $\sim$3 yr. On the other hand, the 
WR 147S 3.6 cm flux density shows a decrease of $\sim$14\% from its 1996 value. Since 
the 3.6 cm flux reported here is the combination of two observing runs for WR 147, we 
can take this value as a highly reliable one. Then, this variation confirms the 
suggestion of WR 147S variability proposed by Williams et al.~(1997) and the variation 
reported previously by Contreras \& Rodr\'\i guez (1999). However, this latter paper 
reports an increase of the radio emission while the present work reports a decrease in 
the same radio flux density. As  Watson et al.~(2002) have suggested for the 
variations in the radio flux at 6 cm, the presence of inhomogeneities in the wind of 
the WR star can cause quite random variations in the radio emission. A precise 
comparison of our 6 cm flux density with the 1995 value (Contreras et al.~1996) cannot 
be made because those data were taken at a lower angular resolution. However, this new 
value is consistent with the 1998 flux reported by Watson et al. (2002) at the same
wavelength. Thus, it may be that flux variations observed at 3.6 cm, which traces 
material inner in the wind, cannot yet be observed at 6 cm, which traces the outer 
regions.

\begin{figure*}[!t]
\begin{center}
\centerline{\includegraphics[width=23pc,angle=270]{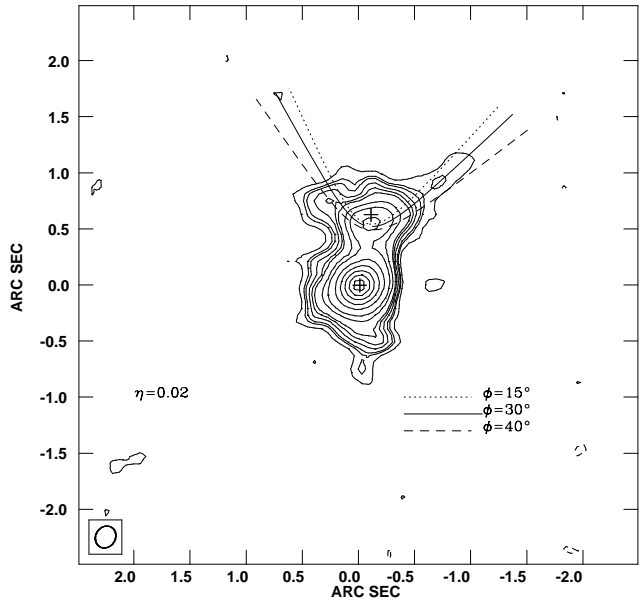}}
\caption{VLA CLEANed map of WR 147 at 3.6 cm. The map was obtained in the same
way as those of Figue 2 with a data weight of ROBUST=-1. We can see two blobs of 
emission to the south of the WR star and a small peak of emission to the E part of 
the shock region, see text. We show a set of fitting curves at different inclination
angles. The best fit curve is that at $30^\circ$.}
\end{center}
\end{figure*}

Morphologically, the high resolution MERLIN observations of Williams et al.~(1997) 
and Watson et al.~(2002) showed that at 6 cm both sources, WR 147S and WR 147N, are 
asymmetric. However, Contreras \& Rodr\'\i guez (1999) found no clear 
elongation in the wind of WR 147S while the northern source was clearly asymmetric. 
With our new VLA observations we still do not find any clear asymmetry in the southern 
source, and again the shocked region (northern component) is obviously elongated. As 
discussed by Contreras \& Rodr\'\i guez (1999), although we confirm that the northern 
source is elongated in the E-W direction, the dimensions derived using MERLIN data
are smaller than ours because they resolve out extended emission. In Table 1 we report 
the deconvolved sizes for both sources at all three observed wavelengths.
At present, it is quite clear that both radio sources show inhomogeneities. Based on 
MERLIN observations at 6 cm, Watson et al.~(2002) have found that both radio components
show radio emission structure. In fact, both emission structures are highly variable 
as in the case of P Cyg's wind, but on a larger time-scale. In our maps we can see a
slightly independent emitting region at the N-E side of the northern source that has 
not been detected in previous VLA observations. Giving the high resolution structure
observed with MERLIN, we suggest that this small region could be a clump entering the 
colliding-wind zone as suggested by Setia Gunawan et al.~(2001). Besides, we detect,
at a 3-4$\sigma$ level, two quite conspicuous condensations in the southern radio 
component located to the South at 3.6 cm (Figure 3). Besides, the clump directly to the 
south seems to be have a counterpart at 6 cm (Figure 2), and additionally in the 6 cm
map we can see another clump in the S-W direction, which does not coincide well with 
one any of the two detected at 3.6 cm. What is most surprising about the 3.6 cm blobs 
is that they seem to lie almost in the same direction and position as those reported 
by Contreras \& Rodr\'\i guez (1999), based on data taken 3 years earlier. Since we do 
not expect them to be the same ones previously observed, then it may be that as
Watson et al.~(2002) have suggested, the WR wind is directed in a north-south 
preferential outflow axis.

Derived spectral indices are shown in Table 2. The northern source possesses a negative 
index indicative of non-thermal emission, which is consistent with previous results.
Based on these values, we can see that both spectral indices for the southern component 
are lower than the expected value for a classical thermal wind. This behaviour can be 
explained if we consider the extreme flux variability at all wavelengths. 


\subsection{Fitting the WR 147 Bow Shock}

We have previously fitted a colliding wind model to the radio morphology
of WR 147N (Contreras \& Rodr\'{\i}guez 1999). Given the improved
quality of our new maps, we now rederive parameters for the system. In
order to apply an analytic model, we will neglect inhomogeneities and
possible mass loss variations, as well as the orbital motion of the
driving stars. The latter assumption is valid because the wind speeds
are much greater than the orbital speeds. A thin shell model for the
wind collision has two fundamental parameters under the assumption of
steady, isotropic winds: the ratio of momentum loss rates $\eta$, and
the angle $\phi$ between the symmetry axis of the collision surface
and the plane of the sky. We adopt the exact solutions of Cant\'o,
Raga \& Wilkin (1996) for the collision surface, and apply the
projection method described by Wilkin (1997) (see also, Raga et
al.~1997). Along the line joining the centers of the two stars, the
winds collide head-on at the stagnation point. If the stars are
separated by a distance $D$, and the stagnation point is at distance
$R_\circ$ from the source of the weaker wind, $\eta$ may be obtained
from the ratio $R_\circ/D = \eta^{1/2}/(1+\eta^{1/2}).$ This ratio
is independent of the angle at which we view the system. In our fit,
we assume that the peak emission of WR 147N corresponds to the stagnation point, 
and measure its position relative to WR 147S, $0\rlap{.}{''}556\pm0\rlap{.}{''}020$ 
The observed separation of the two stars, $0\rlap{.}{''}635\pm0\rlap{.}{''}020$
was taken from Williams et al.~(1997). We obtain $\eta=0.02\pm0.01$. 
This new value of $\eta$ differs from the one obtained by Contreras \&
Rodr\'\i guez (1999) of 0.01, but is still consistent within error. Our revised value
is also in agreement with that used by Dougherty et al.~(2003) to reproduce 
the observed 5 GHz  radio synchrotron and thermal emission. Our assumption 
that the brightest part of the wind collision surface corresponds to the stagnation 
point is intuitive, since the normal shock velocity is greatest at this
location, but is also supported by simulated images of colliding winds
(Raga et al.~1997). Given a value of $\eta$, we then calculated the
projected shock surface for several values of $\phi$ (Fig.~3). Reasonable  
fits to the bow shock wings are obtained for $15^\circ\leq\phi\leq 40^\circ$,
with a best fit of $\phi=30^\circ$. Thus, a lower limit to the inclination
angle of the orbit would be $30^\circ$. Although our $\phi=30^\circ$ curve is
in good agreement with the northwestern wing of the bow shock, the northeastern
wing appears to be more open, suggesting a non-axisymmetric morphology. Although 
analytic models may be generated for non-axisymmetric bow shocks (Wilkin 2000), 
such models introduce too many free parameters to attempt a unique solution 
given the present maps (see Gaensler 2002). However, it would be interesting
to constrain such a model with additional kinematic information derived from 
observations of spatially resolved line emission.

\section{Summary}

All of our observed objects are clear examples of non-classical winds.
P Cyg deviates in at least two ways from the standard model: its wind is anisotropic
and inhomogeneous. Additionally, it is a highly variable source, both in its radio flux 
density at all wavelengths observed and in its morphological structure. This variability 
has been previously reported based on high angular resolution observations. We confirm 
that at a lower resolution we still detect variations. We found that Cyg OB2 
No.~12 is a non-spherical source at 6 cm and we suggest that the observed elongation is 
due to a blob denser than the surrounding wind material. Thus, we report Cyg OB2 No.~12 
as an inhomogeneous wind source for the first time. WR 147 is one of the most 
interesting binary systems. A highly structured wind has been reported at high angular 
resolution. In this work, we detect two possible clumps in the wind of the WR star. 
These blobs are almost in the same direction and position as those reported previously
by Contreras \& Rodr\'\i guez (1999). This confirms the suggestion of Watson et al.~(2002)
of a north-south preferential outflow axis. The shock region (north radio component) 
seems to possess a faint emission peak located to its N-E side which could be a clump 
crossing this zone. We have fitted a theoretical bow shock curve to the northern radio 
component based on our new VLA data. These data give a higher wind momentum ratio, $\eta$, 
than the one derived previously but it is still consistent within error. 
Our new best fitting curve corresponds to an angle $\phi$ of $30^{+10}_{-15}$ degrees 
which gives a lower limit to the inclination angle of the orbit.

 In this way, our new VLA data confirm the non-standard wind behaviour of these three 
interesting sources.

\acknowledgments
We thank Luis F. Rodr\'\i guez for his valuable comments on several versions of the
manuscript. M.E.C, F.P.W. and G.M. thank financial support from DGAPA-PAPIIT and 
CONACyT-Ciencias B\'asicas. F.P.W. also wishes to thank the NSF-International Researchers 
Fellows Program for financial support.


\begin{thebibliography}

\bibitem{} Abbott, D.~C., Bieging, J.~H., \& Churchwell, E.\ 1981, \apj, 250, 645

\bibitem{} Cant\'o, J., Raga, A. \& Wilkin, F.~P.\ 1996, \apj, 469, 729

\bibitem{} Contreras, M.~E., Rodr{\'{\i}}guez, L.~F., Gomez, Y., \& Velazquez, A.\ 
1996, \apj, 469, 329 

\bibitem{} Contreras, M.~E.~\& Rodr{\'{\i}}guez, L.~F.\ 1999, \apj, 515, 762

\bibitem{} Cherepashchuk, A.~M.\ 1990, Soviet Astronomy, 34, 481

\bibitem{} de Jager, C.\ 2001, in P Cyg 2000: 400 Years of Progress, ASP Conf. Proc.,
Vol. 233, eds. M. de Groot and C. Sterken, (San Francisco), 215

\bibitem{} Dougherty, S.~M., Pittard, J.~M., Kasian, L., Coker, R.~F., Williams, 
P.~M., \& Lloyd, H.~M.\ 2003, \aap, 409, 217

\bibitem{} Exter, K.M., Watson, S.K., Barlow, M.J. \& Davis, J.\ 2002, \mnras, 333, 715

\bibitem{} Gaensler, B.~M., Jones, D.~H., \& Stappers, B.~W.\ 2002, \apjl, 580, L137

\bibitem{} Georgiev, L. \& Koenigsberger, G.\ 2002, in ASP Conf. Ser. 260, Interacting
Winds from Massive Stars, eds. A.F.J. Moffat \& N. St-Louis, 393

\bibitem{} Girard, T.~\& Willson, L.~A.\ 1987, \aap, 183, 247

\bibitem{} Luehrs, S.\ 1997, \pasp, 109, 504

\bibitem{} Moffat, A.~F.~J.~\& Robert, C.\ 1994, \apj, 421, 310

\bibitem{} Morris, P. W., van der Hucht, K. A., Crowther, P. A., Hillier, 
D. J., Dessart, L., Williams, P. M. \& Willis, A. J.\ 2000, \aap, 353, 624

\bibitem{} Panagia, N.~\& Felli, M.\ 1975, \aap, 39, 1

\bibitem{} Raga, A.~C., Cant{\' o}, J., Curiel, S., Noriega-Crespo, A., 
\& Raymond, J.~C.\ 1997, RMAA, 33, 157

\bibitem{} Setia Gunawan, D.~Y.~A., de Bruyn, A.G., van der Hucht, K.A. \& 
Williams, P.M.\ 2001, \aap, 368, 484

\bibitem{} Skinner, C.J., Becker, R.H., White, R.L., Exter, M., Barlow, M.J.
\& Davis, R.J. 1998, \mnras, 296, 669

\bibitem{} van Genderen, A.~M.\ 2002, Journal of Astronomical Data, 8, 8

\bibitem{} Watson, S.K., Davis, R.J., Williams, P.M. \& Bode, M.F.\ 2002, \mnras,
334, 631

\bibitem{} Williams, P.~M., Dougherty, S.~M., Davis, R.~J., van der Hucht, K.~A., 
Bode, M.~F., \& Setia Gunawan, D.~Y.~A.\ 1997, \mnras, 289, 1

\bibitem{} Wilkin, F.~P.\ 1997, Ph.D.~Thesis, University of California  

\bibitem{} Wilkin, F.P.\ 2000, \apj, 532, 400  

\bibitem{} Wright, A.~E.~\& Barlow, M.~J.\ 1975, \mnras, 170, 41


\end{thebibliography}
\end{document}